\documentclass[
  aip,
  jcp,
  amsmath,amssymb,
  reprint,
]{revtex4-1}
\usepackage{graphicx,color}
\usepackage{amsmath}
\usepackage{natbib}
\usepackage{epsfig}
\begin{document}

\title{Two-body entropy of two-dimensional fluids}

\author{Boris A. Klumov}

\email{Boris.Klumov@ihed.ras.ru}
\affiliation{Joint Institute for High Temperatures, Russian Academy of Sciences, 125412 Moscow, Russia}

\author{Sergey A. Khrapak}

\email{Sergey.Khrapak@dlr.de}
\affiliation{Joint Institute for High Temperatures, Russian Academy of Sciences, 125412 Moscow, Russia}
\affiliation{Institut f\"ur Materialphysik im Weltraum, Deutsches Zentrum f\"ur Luft- und Raumfahrt (DLR), 82234 We{\ss}ling, Germany}

\date{\today}

\begin{abstract}
The two-body (pair) contribution to the entropy of two-dimensional Yukawa systems is calculated and analyzed. It is demonstrated that in the vicinity of the fluid-solid (freezing) phase transition the pair entropy exhibits an abrupt jump in a narrow temperature range and this can be used to identify the freezing point. Relations to the full excess entropy and some existing freezing indicators are briefly discussed.   
\end{abstract}

\maketitle

The quantity which is often accessible for direct experimental evaluation in the fluid phase is the radial distribution function (RDF) $g(r)$. When interactions are pairwise and are known to a good approximation, important thermodynamic quantities such as pressure and internal  energy can be calculated explicitly as integrals involving $g(r)$.~\cite{HansenBook} Another useful quantity, the two-body contribution to the entropy (or simply two-body entropy), can also be directly calculated from~\cite{BaranyaiPRA1989}
\begin{equation}\label{s2}
s_2=-\frac{n}{2}\int \left[g(r)\ln g(r)-g(r)+1\right]d{\bf r},
\end{equation}      
where $n$ is the particle number density and $s$ is the entropy per particle in units of $k_{\rm B}$. The two-body entropy can be considered as a first non-ideal term in the expansion $s=s_{\rm id}+s_2+s_3+...$, where higher terms involve higher order correlations functions.~\cite{BaranyaiPRA1989}  

From the previous investigations~\cite{BaranyaiPRA1989,LairdPRA1992} it has been known that for conventional dense three dimensional (3D) fluids (not too far from the fluid-solid phase transition) the two-body entropy represents a good approximation for the exact excess entropy $s_{\rm ex}$.~\cite{LairdPRA1992} This has been documented for Lennard-Jones, hard-sphere, inverse-power-law and one-component plasma fluids.~\cite{BaranyaiPRA1989,LairdPRA1992} A similar observation has been reported for a two-dimensional (2D) system of hard discs,~\cite{BorzsakCP1992} although the phase space in the vicinity of crystallization has not been very well resolved. The situation can be different for interaction potentials that result in unusual (anomalous) properties of the respective phase diagram (e.g. Gaussian core model, Hertzian spheres, repulsive shoulder systems), where significant differences between $s_2$ and $s_{\rm ex}$ have been observed.~\cite{GiaquintaCPC2005,FominPRE2010,FominJCP2014} These special cases are not considered here. 

A useful related quantity, the residual multiparticle entropy (RMPE), 
\begin{equation}
\Delta s = s_{\rm ex}-s_2 = \sum_{i=3}^{\infty}s_i,
\end{equation}
can be introduced, where $s_{\rm ex}$ is the excess entropy ($s_{\rm ex}=s-s_{\rm id}$). The RMPE is relatively small in simple dense fluids and is found to vanish in close proximity of the fluid-solid phase transition in 3D.~\cite{Giaquinta1992,GiaquintaPRA1992,SaijaJCP2006} In fact, zero-point of RMPE is a useful indicator of the transition between a disordered (or partially ordered) fluid and a more ordered phase, and this is not restricted to the fluid-solid phase transition.~\cite{SaijaJCP2006} The RMPE-based criterion is applicable to single-component fluids and mixtures in both 2D and 3D.~\cite{SaijaJCP2006,SaijaJCP2000,SaijaJCP2002,GiaquintaCPC2005} However, in 2D the RMPE turns out to vanish somewhat prior to the freezing transition.~\cite{SaijaJCP2000}      

Additionally, it has been recently suggested to use the numerical value of the two-body entropy as an indicator of freezing in 2D.
In particular, it has been observed that $s_2\simeq -4.5\pm 0.5$ at freezing of several 2D systems with different interactions.~\cite{WangJCP2011} Recent experiments and simulations of 2D colloidal hard spheres~\cite{ThorneyworkPRL2015,ThorneyworkJPCM2018} have provided data in support of this proposal. 

Motivated by the success of these freezing indicators and looking for an appropriate comparison with other recently proposed criteria of 2D melting,~\cite{KhrapakJCP2018,KhrapakPRL2019} we performed additional calculations of the behavior of $s_2$ in strongly coupled Yukawa fluids. Our interest to Yukawa fluids is mainly associated with the fact that traditionally the Yukawa (screened Coulomb or Debye-H\"uckel) potential is extensively used as a first approximation to model real interactions between charged particles in colloidal suspensions and (complex) plasma media.~\cite{FortovUFN,FortovPR,FortovBook,KlumovUFN2011,ChaudhuriSM2011} Two-dimensional plasma crystals and fluids constitute an important topic of recent experimental studies.~\cite{NosenkoPRL2004,KryuchkovPRL2018,CouedelUFN2019}  

\begin{figure}
\includegraphics[width=8.0cm]{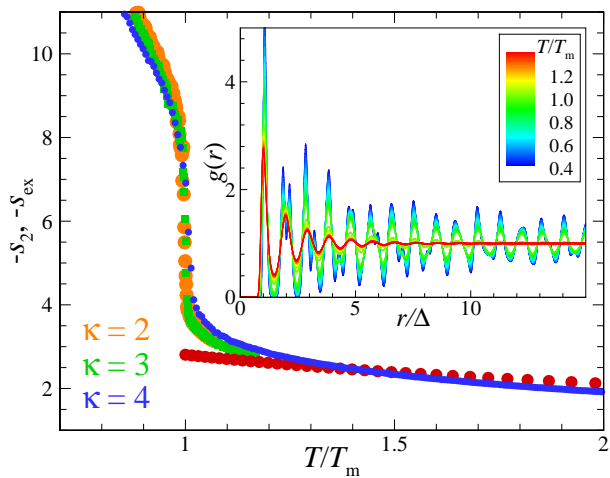}
\caption{Two-body contribution to the excess entropy, $s_2$, of 2D Yukawa systems versus the reduced temperature $T/T_{\rm m}$, where $T_{\rm m}$ is the melting temperature. Results for three different $\kappa$ are shown (see the legend). Red dots correspond to the ``exact'' excess entropy, $s_{\rm ex}$, calculated from the thermodynamic consideration. The inset shows the RDFs used to evaluate $s_2$ from Eq.~(\ref{s2}).}
\label{Fig1}
\end{figure}

\begin{figure}
\includegraphics[width=8.0cm]{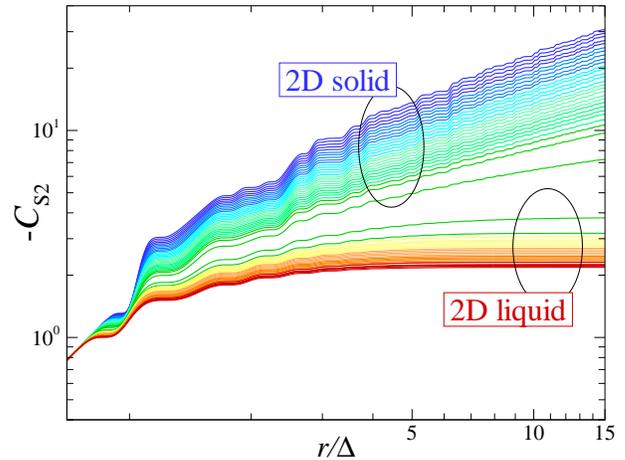}
\caption{Cumulative two-body entropy $C_{s2}$ versus reduced distance $r/\Delta$ for a 2D Yukawa system with $\kappa=3$. The curves from top to bottom differ by a uniform increase in temperature of $\simeq 0.01 T_{\rm m}$. Two well separated branches are labeled as 2D solid and 2D liquid (see the text for details).}
\label{Fig2}
\end{figure}

We have generated RFDs, necessary to calculate $s_2$, by performing molecular dynamics (MD) simulations using the LAMMPS package.~\cite{LAMMPS} The system of $4 \times 10^4$ particles has been simulated in the Nose-Hoover $NVT$ ensemble with periodic boundary conditions. Starting from equilibrium crystal at $T\simeq 0.1T_{\rm m}$ the system has been heated up to $T\simeq 2T_{\rm m} $ using a constant temperature step of $\simeq 0.01T_{\rm m}$, where $T_{\rm m}$ denotes the melting temperature. Each configuration has been equilibrated during $10^6$ time steps. Three different Yukawa systems, characterized by different screening parameters $\kappa=2,$ 3, and 4, have been considered. Here the screening parameter, $\kappa=\Delta/\lambda$, is the ratio of the characteristic inter-particle separation $\Delta=1/\sqrt{n}$ to the screening length $\lambda$. 
The main results are summarized in Figs.~\ref{Fig1} and \ref{Fig2}. 
 
Figure~\ref{Fig1} shows the dependence of $s_2$ on the reduced temperature for three systems considered. Two different scalings are clearly observed. First, $-s_2$ smoothly increases on approaching the boundary of the fluid phase stability.  
In the vicinity of the melting temperature, $-s_2$ exhibits an abrupt jump to much higher values. The observed jump corresponds to a rather narrow temperature range, which can be used to identify $T_{\rm m}$. The dependence of $-s_2$ on $T/T_{\rm m}$ appears quasi-universal for the considered systems. Such a strong inclination of $s_2$ on approaching the fluid-solid phase transition (also seen in Fig. 4 of Ref.~\onlinecite{ThorneyworkJPCM2018}) makes a particular value of $s_2$ impractical in determining the location of the phase change. We observe from Fig.~\ref{Fig1} that the condition $s_2\simeq -4.5$ would indeed allow us to locate the freezing point. However, it also shows that any other number from the range $4\lesssim - s_2 \lesssim 8$ would provide essentially the same accuracy in the considered situation. 

In Figure~\ref{Fig2} we plot the cumulative two-body entropy calculated from (in 2D geometry)
\begin{equation}
C_{s2}(R)=-\pi\int_0^{R} \left[g(x)\ln g(x)-g(x)+1\right]xd{x},
\end{equation}
where $x=r/\Delta$ is the reduced distance and $R$ represents the upper integration limit [in passing we note that the points shown in Fig.~\ref{Fig1} correspond to $C_{s2}(15)$]. Two branches of curves can be clearly identified. The lower branch is characterized by a relatively fast convergence and is identified as the fluid branch. The upper branch seems to diverge and apparently corresponds to the solid phase. In terms of $s_2$, the fluid branch ends near $s_2\simeq -4$, in reasonable agreement with the original prediction.~\cite{WangJCP2011} However, the phase indicator based on the cumulative two-body entropy $C_{s2}$ seems more advantageous than that based on the value of $s_2$ alone.            

The last important point concerns the relation between $s_2$ and the ``exact'' excess entropy $s_{\rm ex}$. The latter exhibits a quasi-universal dependence on $T/T_{\rm m}$, which can be regarded as a 2D analogue of the Rosenfeld-Tarazona scaling.~\cite{RosenfeldMolPhys1998,KhrapakPoP2015,SemenovPoP2015,KryuchkovJCP2017} We have calculated the excess entropy of Yukawa fluid with $\kappa\simeq 3.5$ ($1/\sqrt{\pi n \lambda^2}=2$) using accurate thermodynamic data from Ref.~\onlinecite{KryuchkovJCP2017}. The results are shown in Fig.~\ref{Fig1} by red dots. We observe that $s_2$ and $s_{\rm ex}$ are relatively close in the parameter regime investigated, except in close proximity of the fluid-solid phase transition. This is in striking contrast with usual simple 3D fluids, where the agreement between $s_2$ and $s_{\rm ex}$ is particularly good near the freezing point,~\cite{BaranyaiPRA1989,LairdPRA1992} and the RMPE vanishes there.~\cite{Giaquinta1992,GiaquintaPRA1992,SaijaJCP2006} Our present observation is in qualitative agreement with previous results on 2D Lennard-Jones fluid which demonstrated that $\Delta s =0$ occurs at densities systematically lower than the freezing-point densities.~\cite{SaijaJCP2000}  

Can there be some additional special peculiarities about Yukawa systems in this context? This is not very likely. The freezing point excess entropy of Yukawa fluid with $\kappa\simeq 3.5$, $s_{\rm ex}\simeq -2.8$, is comparable to that of other 2D fluids with soft repulsive interactions. For instance, we have obtained $s_{\rm ex}\simeq -3.3$ at freezing of a 2D one-component plasma with logarithmic repulsion,~\cite{KhrapakCPP2016,OCPLog}   $s_{\rm ex}\simeq -3.1$ at freezing of a 2D one-component plasma with Coulomb ($\propto 1/r$) repulsion,~\cite{KhrapakCPP2016} and  $s_{\rm ex}\simeq -2.9$ at freezing of a 2D system with isotropic dipole-like ($\propto 1/r^3$) repulsion.~\cite{KhrapakPRE2018} Finally, using a simple equation of state for hard discs,~\cite{HendersonMolPhys1975} we have estimated  $s_{\rm ex}\simeq -3.5$ at freezing of a hard disc fluid. Note that all these values are considerably higher than the expected $s_2\sim -4.5$ near the freezing point. This indicates that $s_2$ and $s_{\rm ex}$ are apparently not very close in 2D fluids near solidification (where $s_2$ underestimates $s_{\rm ex}$) and that the RMPE vanishes somewhat prior to the freezing transition in 2D.             
 
We would like to thank M. Schwabe and H. Thomas for careful reading of the manuscript. 
The work leading to this publication was partly supported by the German Academic Exchange Service (DAAD) with funds from the German Aerospace Center (DLR). 

\bibliographystyle{aipnum4-1}
\bibliography{S2_References}

\end{document}